\documentclass[preprint,showpacs,preprintnumbers,amsmath,amssymb]{revtex4}

\usepackage{graphicx}
\usepackage{dcolumn}
\usepackage{bm}
\usepackage{quantum}

\def\sigm{\op{\sigma_-}}
\def\sigp{\op{\sigma_+}}
\def\aout{\a_{out}}

\def\eout{\e_{out}}
\def\ain{\a_{in}}

\def\domega{\Delta\omega}

\def\aindag{\a_{in}^{\dagger}}

\def\Dalpha[#1]{D_\alpha\left(#1\right)}

\def\signzab{\op{\sigma_z^{12}}}
\def\signmab{\op{\sigma_-^{12}}}

\def\signzbc{\op{\sigma_z^{23}}}
\def\signmbc{\op{\sigma_-^{23}}}

\def\apump{\a_{\nu+\Delta}}
\def\adagpump{\adag_{\nu+\Delta}}
\def\bpump{\b_{\nu+\Delta}}
\def\bdagpump{\bdag_{\nu+\Delta}}

\def\S{\hat{\op{S}}}

\begin{document}

\preprint{APS/123-QED}

\title{Dispersive properties and giant Kerr non-linearities in Dipole Induced Transparency}

\author{Edo Waks}
\affiliation{
E.L. Ginzton Labs\\
Stanford University, Stanford, CA, 94305
}%

\author{Jelena Vuckovic}
\affiliation{
E.L. Ginzton Labs\\
Stanford University, Stanford, CA, 94305
}%

\date{\today}

\begin{abstract}
We calculate the dispersive properties of the reflected field from
a cavity coupled to a single dipole. We show that when a field is
resonant with the dipole it experiences a $\pi$ phase shift
relative to reflection from a bare cavity if the Purcell factor
exceeds the bare cavity reflectivity. We then show that optically
Stark shifting the dipole with a second field can be used to
achieve giant Kerr non-linearites. It is shown that currently
achievable cavity lifetimes and cavity quality factors can allow a
single emitter in the cavity to impose a nonlinear $\pi$ phase
shift at the single photon level.
\end{abstract}

\pacs{Valid PACS appear here}
\maketitle


Optical non-linearities play an important role in quantum optics,
quantum information processing, and also in design of practical
quantum electronic devices.  One of the most commonly addressed
non-linearities is the Kerr effect which can result in cross-phase
modulation and two-photon absorption.  These effects have a large
number of applications for optical detection, all-optical
switching, and quantum
computation~\cite{ImotoHaus85,HarrisYamamoto98}. The main
difficulty in achieving these applications is that conventional
materials offer only a very small non-linear response, which is
significantly outweighed by linear absorption. Furthermore,
applications in quantum optics and quantum information often
require that these non-linearities be created by a small number of
photons, or sometimes even a single photon. There are very few
situations where one can even approach this regime.

One of the few cases where Kerr non-linearities with a small
number of photons can be observed is in atomic vapors using
Electromagnetically Induced Transparency
(EIT)~\cite{HarrisField90}. Due to the large atomic coherence of
these systems, cross phase modulation and two-photon absorption
can be observed with only a small number of
photons~\cite{BrajeBalic04}. The central idea for these schemes
was originally presented by Schmidt and
Imamoglu~\cite{SchmidtImamoglu96}.  This scheme exploits EIT in a
4-state atom where the interaction between two weak pulses is
mediated by a strong coupling laser. A limitations of the
Schmidt-Imamoglu proposal is that large group velocity mismatch
between the two interacting pulses puts a fundamental lower limit
on the required intensity for a full $\pi$ phase shift to a few
photons per cubic wavelength~\cite{HarrisHau99}.

In this paper we explore the dispersive properties of light that
is reflected from a cavity containing a single dipole emitter. We
calculate absorptive and dispersive properties of the reflection
coefficient as a function of the coupling strength between the
dipole and cavity.  We show that for large values of the Purcell
factor, losses due to cavity leakage and dipole absorption are
cancelled.  This is a manifestation of destructive interference
which inhibits the light from entering the cavity, and is known as
Dipole Induced Transparency (DIT)~\cite{WaksVuckovic05}.  At the
same time, when the Purcell factor exceeds the bare cavity
reflectivity, the presence of a dipole imposes a $0$ phase shift
on the reflected field that is resonant with the dipole frequency,
whereas a bare cavity would impose a $\pi$ phase shift. This
change of phase has been previously studied in the strong coupling
regime of cavity QED~\cite{DuanKimble04}.  Our result shows that
only weak coupling is required to observe this dipole induced
phase shift. Thus, DIT provides a special condition where we can
drive the dipole on resonance and create large phase shifts while
not suffering from absorption.

The large dipole induced phase shifts in DIT are sharply peaked
near the resonant frequency of the emitter, creating a highly
dispersive region with large group delays.  This large dispersion
allows us to create giant Kerr non-linearities by optically Stark
shifting the dipole. This method of achieving non-linearities is
significantly easier to implement than the Schmidt-Imamoglu scheme
using EIT. Instead of requiring a 4-level system, we only require
a 3-level system for the dipole. When using a quantum dot emitter,
these levels can be readily provided by the single exciton and
bi-excitonic resonance. Furthermore, we do not need an external
coupling laser, since coupling is provided by the vacuum Rabi
frequency of the dipole system.  Finally, our proposal can exploit
the large Rabi frequencies provided by photonic crystal
micro-cavities to achieve non-linearities with only a single
emitter.  We show that with presently achievable cavity lifetimes
and vacuum Rabi frequencies, working only in the weak coupling
regime, we can achieve a full non-linear $\pi$ phase shift using
only a single emitter and a single photon in the cavity.

\begin{figure}
\centering\includegraphics[width=7cm]{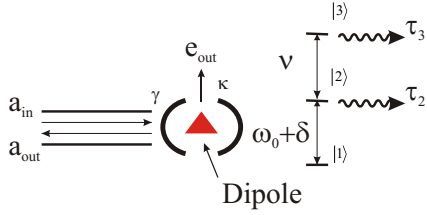} \caption{Setup
for large Kerr non-linearities off of reflection from a cavity
coupled to a single dipole.} \label{fig:setup}
\end{figure}

The system we study in this paper is shown in
Fig.~\ref{fig:setup}.  An external waveguide field is reflected
off of a single-sided cavity containing a dipole.  We define
$\gamma$ as the energy coupling rate from cavity to waveguide,
while $\kappa$ is the energy coupling rate into parasitic leaky
modes.  The dipole has three states, denoted $\|1>$, $\|2>$, and
$\|3>$, with transition frequencies $\omega_0+\delta$ and $\nu$,
where $\omega_0$ denotes the central frequency of the cavity. We
assume that $\delta,\omega_0-\nu<\gamma+\kappa$, so that both
transitions can couple to the cavity mode.  We denote $\signmab$
and $\signmbc$ as the lowering operators for the dipole, and $\b$
as the bosonic annihilation operator for the cavity field.

The Hamiltonian of the system is given by
  \begin{equation}
    H = H_s + \sum_{n=1}^N\hbar \left[ g_1\bdag\signmab  +
    g_2\left(\b+\bdag\right)\signmbc + \mbox{H.c.}  \right]
  \end{equation}
where $g_1$ and $g_2$ are the vacuum rabi frequencies for the 1-2
and 2-3 transitions respectively.  $H_s$ is the Hamiltonian of the
uncoupled systems and the cavity-waveguide interaction terms.  In
the first part of the paper we only consider the 1-2 transition to
calculate the dispersive properties of the cavity in the presence
of a dipole.  Thus, we set $g_2=0$.  In the second part of the
paper we add the 2-3 transition and consider the case where the
dipole population is mainly in state $\|1>$.  Thus, the 2-3
transition does not drive the cavity but can still create an
optical Stark shift on state $\|2>$. In order to properly derive
the Stark shift term, we do not yet make the rotating wave
approximation for the interaction between the cavity and
$\signmbc$.

Using the above Hamiltonian, along with standard cavity
input-output formalism~\cite{WallsMilburn}, the Heisenberg picture
equations of motion for the operators are given by
  \begin{eqnarray}
    \frac{d\b}{dt} & = & \left(-i\omega_0 + \gamma/2 + \kappa/2 \right) \b -
    \sqrt{\gamma}\aout  - \sqrt{\kappa}\eout -
    ig_1\signmab  \label{eq:Heisenbergb} \\
    \frac{d\signmab}{dt} & = & \left(-i\left(\omega_0+\delta\right) +  \frac{\tau_2}{2}\right)\signmab -ig_1\b -ig_2\left(\b+\bdag\right)\signmab\signmbc+\f
    \label{eq:HeisenbergSig12} \\
    \frac{d\signmbc}{dt} & = & \left(-i\nu + \frac{\tau_3}{2}
    \right) \signmbc - ig_2\left(\b+\bdag\right)\signzbc + \h \label{eq:HeisenbergSig23}
  \end{eqnarray}
The operators $\f$ and $\h$ are noise operators that are needed to
conserve the commutation relation of the dipole operators.  The
above equations are derived with the assumption that there is
virtually no population in the states $\|2>$ and $\|3>$, so
$\signzab$ and $\signzbc$ are time invariant, and $\signmbc$ does
not drive the cavity mode. The condition for this assumption to be
valid is given by $\langle\sigp\sigm\rangle\ll 1$, which is
equivalent to the condition $\langle\aindag\ain\rangle\ll
g_1^2/\gamma$ for an input field that is resonant with the
dipole~\cite{WaksVuckovic05}. This condition is well satisfied in
the operating regime we consider.  The input and output fields are
related by
 \begin{equation}
    \aout-\ain  =  \sqrt{\gamma}\b \label{eq:ascat}
  \end{equation}

We begin by first studying the dispersive properties of reflected
light when we only have the transition $\signmab$ and set $g_2=0$.
In this case, if we apply an input field at frequency $\omega$,
the reflection coefficient can be solved using
Eq.~\ref{eq:Heisenbergb}-\ref{eq:ascat} and is given by
  \begin{equation}
    r(\omega) = \frac{i\domega +
    \frac{g_1^2}{i(\domega+\delta) + \tau_2/2}-\gamma/2 +\kappa/2}{i\domega+ \frac{g_1^2}{i(\domega+\delta) + \tau_2/2}+\gamma/2 +\kappa/2}
  \end{equation}
We define the amplitude and phase of the reflection coefficient by
$r = \sqrt{R}e^{i\Phi_r}$, where $R$ is the reflectivity of the
cavity and $\Phi_r$ is the phase shift imposed on the reflected
field.

\begin{figure}
\centering\includegraphics[width=7cm]{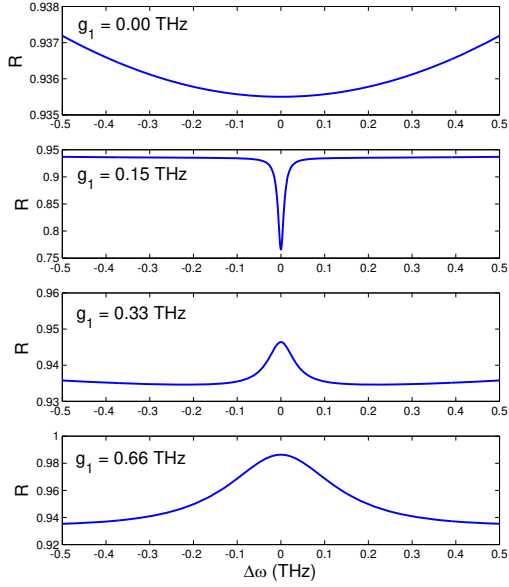}
\caption{Cavity reflection coefficient for different values of
$g$.} \label{fig:reflection}
\end{figure}

Fig.~\ref{fig:reflection} plots the cavity reflectivity as a
function of detuning from the cavity resonance for several values
of $g_1$. We set $\gamma=6THz$ and $\kappa=0.1THz$, which is the
decay rate of a cavity with a quality factor of $Q=10,000$ (here Q
is defined as $Q=\omega_0/\kappa$). The dipole is assumed to be
resonant with the center frequency of the cavity so that
$\delta=0$, while $\tau_2=1GHz$, a value taken from experimental
measurements on quantum dots~\cite{VuckovicFattal03}.

When $g_1=0$ the cavity is not perfectly reflecting due to the
coupling to leaky modes given  by $\kappa$.  Introducing a small
$g_1$ increases the cavity loss on-resonance, because the dipole
behaves as an absorber.  However, when $g_1$ is increased to
higher values, the cavity reflection improves. This can be
understood from the reflection coefficient at $\domega=0$, which
is given by
  \begin{equation}\label{eq:ResRef}
    r(\omega_0)=(F_p-r_0)/(F_p+1)
  \end{equation}
where $F_p=4g^2/[\tau_2(\gamma+\kappa)]$ is known as the Purcell
factor, and $r_0 = (\gamma-\kappa)/(\gamma+\kappa)$ is the
reflection coefficient for a bare cavity with no dipole. When
$F_p\approx r_0$, the cavity is very lossy because most of the
field is absorbed by the dipole. However, when $F_p\gg r_0$ the
cavity becomes very reflective. This is the result of destructive
interference of the cavity mode when it is coupled to the dipole,
which prevents light from entering the cavity and therefore
inhibits losses~\cite{WaksVuckovic05}.

\begin{figure}
\centering\includegraphics[width=7cm]{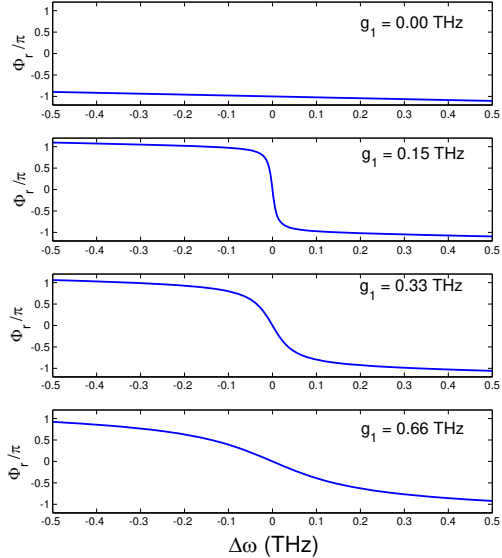}
\caption{Phase shift after reflection from cavity for several
values of $g$.} \label{fig:phaseshift}
\end{figure}

We next plot $\Phi_r$, normalized by $\pi$, in
Fig.~\ref{fig:phaseshift} for different values of $g_1$. When
$g_1=0$ the field experiences a $\pi$ phase shift on reflection,
which is the expected behavior for reflection off of a bare
cavity. For nonzero $g_1$, we see significant modification of the
reflection phase near the dipole resonance. Instead of a $\pi$
phase shift, we now have a $0$ phase shift. This shift quickly
changes to $\pi$ at a detuning of $\domega\approx
g_1^2/(\gamma/2+\kappa/2)$.  This phase change can be easily
understood from Eq.~\ref{eq:ResRef}, which shows that when
$F_p>r_0$ the reflection coefficient is positive, while $F_p<r_0$
implies a negative reflection coefficient.

It is significant to note that the dipole imposes a large change
of phase whenever $F_p>r_0$. In previous work by Duan and
Kimble~\cite{DuanKimble04}, it has been shown that this change in
reflection phase can be used to perform a controlled phase gate by
performing successive reflections of two photons off of a cavity
mode.  In that proposal, the authors speculated that the strong
coupling regime is required to achieve proper performance. Here we
shows that strong coupling is not required for implementation of
the Duan-Kimble proposal. A Purcell factor greater than $r_0$
already achieves this desired phase shift. We need only create
sufficiently large values for $F_p$ so that the cavity is not
lossy.

The sharp dispersive feature of the reflection coefficient also
opens up the possibility to achieve large Kerr non-linearity. If
we can shift the dispersion curve by a very small amount, on the
order of $\domega\approx g_1^2/(\gamma/2+\kappa/2)$, we can change
the reflection phase from $\pi$ to $0$.  The shift in dispersion
can be created by optically Stark shifting the $1$-$2$ transition
by applying an off-resonant field on the $2$-$3$ transition.

To calculate the optical Stark shift, we assume the input field
has two frequency components, one at $\omega$ and the other at
$\nu+\Delta$.  The component at $\nu+\Delta$ is the field
responsible for creating a Stark shift.  The response of the
cavity at this frequency is given by substituting $\ain = \apump
e^{-i(\nu+\Delta)t}$ and taking the Fourier transform of
Eq.~\ref{eq:Heisenbergb} at frequency $\nu+\Delta$. This gives us
  \begin{equation}
    \bpump =
    \frac{-\sqrt{\gamma}\apump}{i(\omega_0-\nu-\Delta)+\gamma}
    \approx\frac{-\apump}{\sqrt{\gamma}}
  \end{equation}
We now assume that $\signmbc$ is driven mainly by the field
component at frequency $\nu+\Delta$.  Using this approximation, we
calculate this Fourier component of Eq.~\ref{eq:HeisenbergSig23},
which is given by
  \begin{equation}
    \signmbc(\nu+\Delta) =
    \frac{-ig_2\signzbc\left(\bpump + \bdagpump\right)}{i\Delta +
    \frac{\tau_3}{2}}
  \end{equation}
We substitute the above expression back into
Eq.~\ref{eq:HeisenbergSig12} and make the rotating wave
approximation, which gives us
  \begin{equation}
    \frac{d\signmab}{dt}  =  \left(-i\left(\omega_0+\delta\right) +
    \frac{\tau_2}{2} - i\S \right)\signmab + ig_1\signzab\b + \sqrt{\tau_3}\f \label{eq:HeisenbergStark}
  \end{equation}
where
  \begin{equation}
    \S =
    \frac{i2g_2^2\bdagpump\bpump}{i\Delta
    + \frac{\tau_3}{2}}
  \end{equation}
The Stark operator $\S$ has a real and imaginary component. The
real component gives the optical Stark shift, while the imaginary
component represents loss due to two-photon absorption.  When
$\Delta\gg\tau_3$ this operator represents an energy shift that is
proportional to the number of photons at frequency $\nu+\Delta$.
If, instead, the field is on resonance with the transition from
$\|2>$ to $\|3>$ so that $\Delta\to 0$, the Stark operator becomes
a loss coefficient. When pumped by a bright field $\apump$, we can
substitute $\bdagpump\b=\langle\adagpump\apump\rangle/\gamma$.

Using the above expressions, we can once again solve for the
cavity reflection coefficient when the cavity is driven by a
monochromatic input at frequency $\omega$. The reflection
coefficient is now given by
  \begin{equation}
    r(\omega) = \frac{i\domega +
    \frac{g_1^2}{i(\domega+\delta+\S) + \tau_2} -\gamma/2 +\kappa/2 }{i\domega+ \frac{g_1^2}{i(\domega+\delta+\S) + \tau_2}+\gamma/2 +\kappa/2}
  \end{equation}
which is nearly identical to the $g_2=0$ case, except that the
dipole is now detuned by an additional value determined by $\S$.

\begin{figure}
\centering\includegraphics[width=7cm]{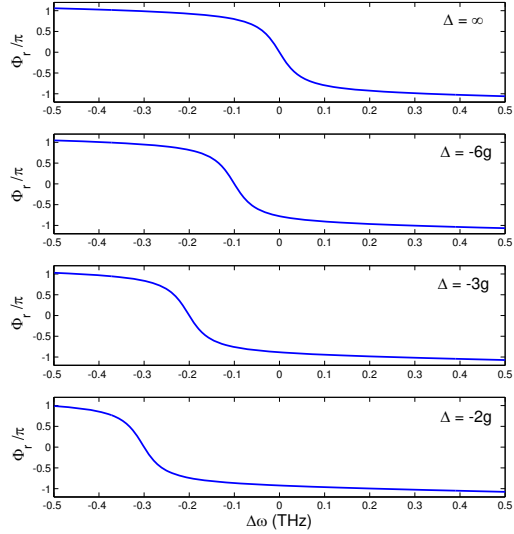}
\caption{Reflection phase shift in presence of a second field at
frequency $\nu+\Delta$ for several values of $\Delta$.  We set
$g=0.3THz$.} \label{fig:kerrshift}
\end{figure}

We consider the case where there is exactly one photon in the
cavity at frequency $\nu+\Delta$, in which case we can substitute
$\bdagpump\bpump=1$, and $\Delta\gg\tau_3$ so that the Stark
operator causes only a phase shift. To calculate the Stark shift
we must have a value for $g_2$.  In the case of a quantum dot
where we use the single and bi-exciton transitions as the two
excited states, it is reasonable to assume that $g_2=g_1=g$.  This
is because both transitions represent an absorption of a photon by
an exciton in the quantum dot. For other systems, the value of
$g_2$ must be measured or calculated from the matrix element
between $\|2>$ and $\|3>$.

Making the assumption that $g_1=g_2=g$, Fig.~\ref{fig:kerrshift}
plots the phase of the reflection coefficient for several values
of $\Delta$, with $g=0.3THz$.  We first plot the case of
$\Delta=\infty$, in which case there is no optical Start shift.
The curve is therefore identical to the one in
Fig.~\ref{fig:phaseshift}.  As we bring the field closer to
resonance with the second transition of the dipole, the resonant
frequency of the first transition is shifted by $2g^2/\Delta$. The
phase shift curve will therefore be translated to the new resonant
frequency of the dipole.  As the figure shows, $\Delta\approx -6g$
is enough to change the reflection phase from $0$ to $\pi$ in the
presence of a single photon.  In general, when
$|\Delta|=\kappa/2+\gamma/2$ we have a frequency shift of
$2g^2/(\kappa+\gamma)$, which is sufficiently large to change the
phase by $\pi$.  This means that any field which is resonant with
the cavity can provide a $\pi$ phase shift.

The system we consider exhibits large Kerr non-linearities due to
two properties.  The first is that the confinement of the cavity
field creates large values of $g$, which in turn generate very
strong optical Stark shifts.  But there is a second, more critical
property.  The largest phase difference is experienced near
resonance with the dipole.  In a normal system, we would not be
able to drive a dipole resonantly without simultaneously suffering
from large losses.  In the case of a DIT, however, we can drive
the system on resonance and not suffer from absorption. As
Fig.~\ref{fig:reflection} shows, the losses are smallest when the
input field is resonant with the cavity.  This is because when we
drive the system on resonance, the cavity field destructively
interferes and thus inhibits any external driving field from
entering the cavity.  This eliminates absorption, allowing us to
work in a regime where we can go on resonance while not suffering
from optical losses.

In conclusion, we have calculated the dispersive properties of
reflection from a cavity seeded with a single emitted, and shown
than when $F_p>r_0$, where $r_0$ is the bare cavity reflectivity,
a $0$ phase shift is induced on the reflected field, instead of a
$\pi$ phase shift for a bare cavity. We showed that by optically
Stark shifting the dipole, we can create large cross-phase
modulation angles at the single photon level.

This work was funded in part by the MURI center for photonic
quantum information systems (ARO/DTO Program DAAD19-03-1-0199),
and a Department of Central Intelligence postdoctoral grant.


\begin{thebibliography}{10}
\expandafter\ifx\csname
natexlab\endcsname\relax\def\natexlab#1{#1}\fi
\expandafter\ifx\csname bibnamefont\endcsname\relax
  \def\bibnamefont#1{#1}\fi
\expandafter\ifx\csname bibfnamefont\endcsname\relax
  \def\bibfnamefont#1{#1}\fi
\expandafter\ifx\csname citenamefont\endcsname\relax
  \def\citenamefont#1{#1}\fi
\expandafter\ifx\csname url\endcsname\relax
  \def\url#1{\texttt{#1}}\fi
\expandafter\ifx\csname
urlprefix\endcsname\relax\def\urlprefix{URL }\fi
\providecommand{\bibinfo}[2]{#2}
\providecommand{\eprint}[2][]{\url{#2}}

\bibitem[{\citenamefont{Imoto et~al.}(1985)\citenamefont{Imoto, Haus, and
  Yamamoto}}]{ImotoHaus85}
\bibinfo{author}{\bibfnamefont{N.}~\bibnamefont{Imoto}},
  \bibinfo{author}{\bibfnamefont{H.~A.} \bibnamefont{Haus}}, \bibnamefont{and}
  \bibinfo{author}{\bibfnamefont{Y.}~\bibnamefont{Yamamoto}},
  \bibinfo{journal}{Phys. Rev. A} \textbf{\bibinfo{volume}{32}},
  \bibinfo{pages}{2287} (\bibinfo{year}{1985}).

\bibitem[{\citenamefont{Harris and Yamamoto}(1998)}]{HarrisYamamoto98}
\bibinfo{author}{\bibfnamefont{S.~E.} \bibnamefont{Harris}} \bibnamefont{and}
  \bibinfo{author}{\bibfnamefont{Y.}~\bibnamefont{Yamamoto}},
  \bibinfo{journal}{Phys. Rev. Lett.} \textbf{\bibinfo{volume}{81}},
  \bibinfo{pages}{3611} (\bibinfo{year}{1998}).

\bibitem[{\citenamefont{Harris et~al.}(1990)\citenamefont{Harris, Field, and
  Imamoglu}}]{HarrisField90}
\bibinfo{author}{\bibfnamefont{S.~E.} \bibnamefont{Harris}},
  \bibinfo{author}{\bibfnamefont{J.~E.} \bibnamefont{Field}}, \bibnamefont{and}
  \bibinfo{author}{\bibfnamefont{A.}~\bibnamefont{Imamoglu}},
  \bibinfo{journal}{Phy. Rev. Lett.} \textbf{\bibinfo{volume}{64}},
  \bibinfo{pages}{1107} (\bibinfo{year}{1990}).

\bibitem[{\citenamefont{Braje et~al.}(2004)}]{BrajeBalic04}
\bibinfo{author}{\bibfnamefont{D.~A.} \bibnamefont{Braje}}
  \bibnamefont{et~al.}, \bibinfo{journal}{Phys. Rev. Lett.}
  \textbf{\bibinfo{volume}{93}}, \bibinfo{pages}{183601}
  (\bibinfo{year}{2004}).

\bibitem[{\citenamefont{Schmidt and Imamoglu}(1996)}]{SchmidtImamoglu96}
\bibinfo{author}{\bibfnamefont{H.}~\bibnamefont{Schmidt}} \bibnamefont{and}
  \bibinfo{author}{\bibfnamefont{A.}~\bibnamefont{Imamoglu}},
  \bibinfo{journal}{Opt. Lett.} \textbf{\bibinfo{volume}{21}},
  \bibinfo{pages}{1936} (\bibinfo{year}{1996}).

\bibitem[{\citenamefont{Harris and Hau}(1999)}]{HarrisHau99}
\bibinfo{author}{\bibfnamefont{S.~E.} \bibnamefont{Harris}} \bibnamefont{and}
  \bibinfo{author}{\bibfnamefont{L.~V.} \bibnamefont{Hau}},
  \bibinfo{journal}{Phys. Rev. Lett.} \textbf{\bibinfo{volume}{82}},
  \bibinfo{pages}{4611} (\bibinfo{year}{1999}).

\bibitem[{\citenamefont{Waks and Vuckovic}()}]{WaksVuckovic05}
\bibinfo{author}{\bibfnamefont{E.}~\bibnamefont{Waks}} \bibnamefont{and}
  \bibinfo{author}{\bibfnamefont{J.}~\bibnamefont{Vuckovic}},
  \bibinfo{note}{quant-ph/0510228}.

\bibitem[{\citenamefont{Walls and Milburn}(1994)}]{WallsMilburn}
\bibinfo{author}{\bibfnamefont{D.}~\bibnamefont{Walls}} \bibnamefont{and}
  \bibinfo{author}{\bibfnamefont{G.}~\bibnamefont{Milburn}},
  \emph{\bibinfo{title}{Quantum Optics}} (\bibinfo{publisher}{Springer},
  \bibinfo{address}{Berlin}, \bibinfo{year}{1994}).

\bibitem[{\citenamefont{Vuckovic et~al.}(2003)}]{VuckovicFattal03}
\bibinfo{author}{\bibfnamefont{J.}~\bibnamefont{Vuckovic}}
  \bibnamefont{et~al.}, \bibinfo{journal}{App. Phys. Lett.}
  \textbf{\bibinfo{volume}{82}}, \bibinfo{pages}{3596} (\bibinfo{year}{2003}).

\bibitem[{\citenamefont{Duan and Kimble}(2004)}]{DuanKimble04}
\bibinfo{author}{\bibfnamefont{L.~M.} \bibnamefont{Duan}} \bibnamefont{and}
  \bibinfo{author}{\bibfnamefont{H.~J.} \bibnamefont{Kimble}},
  \bibinfo{journal}{Phy. Rev. Lett.} \textbf{\bibinfo{volume}{92}},
  \bibinfo{pages}{127902} (\bibinfo{year}{2004}).

\end{thebibliography}
\end{document}